\documentclass[journal]{IEEEtran}
\usepackage{amsmath,amsfonts}
\usepackage{algorithmic}
\usepackage{algorithm}
\usepackage{array}
\usepackage[caption=false,font=normalsize,labelfont=sf,textfont=sf]{subfig}
\usepackage{textcomp}
\usepackage{stfloats}
\usepackage{url}
\usepackage{verbatim}
\usepackage{graphicx}
\usepackage{cite}
\graphicspath{{./Figures/}}

\begin{document}

\title{A Fast Power Spectrum Sensing Solution for Generalized Coprime Sampling}

\author{Kaili Jiang, Dechang Wang, Kailun Tian, Hancong Feng, Yuxin Zhao, Junyu Yuan and Bin Tang 

\thanks{Manuscript received November 23, 2023; revised month date, year; accepted month date, year. Date of publication month date, year; date of current version month date, year. This work was supported in part by the National Natural Science Foundation of China under Grant 62301119, and in part by the Key Project of the National Defense Science and Technology Foundation Strengthening Plan 173 under Grand 2022-JCJQ-ZD-010-12. The associate editor coordinating the review of this manuscript and approving it for publication was xxx. (Corresponding author: Kaili Jiang).}

\thanks{Kaili Jiang, Dechang Wang, Kailun Tian, Hancong Feng, Yuxin Zhao, Junyu Yuan and Bin Tang are with the School of Information and Communication Engineering, University of Electronic Science and Technology of China, Chengdu, Sichuan, 611731, China. (e-mail: jiangkelly@uestc.edu.cn, c13844033835@163.com, kailun$\_$tian@163.com, 2927282941@qq.com, 105 1172535@qq.com, Jyyuan@uestc.edu.cn, bint@uestc.edu.cn).}

\thanks{Digital Object Identifier}}

\markboth{Journal of \LaTeX\ Class Files,~Vol.~xx, No.~x, Month~Year}%
{Jiang \MakeLowercase{\textit{et al.}}: A Fast Power Spectrum Sensing Solution for Generalized Coprime Sampling}


\maketitle

\begin{abstract}
The growing scarcity of spectrum resources, wideband spectrum sensing is required to process a prohibitive volume of data at a high sampling rate. For some applications, spectrum estimation only requires second-order statistics. In this case, a fast power spectrum sensing solution is proposed based on the generalized coprime sampling. By exploring the sensing vector inherent structure, the autocorrelation sequence of inputs can be reconstructed from sub-Nyquist samples by only utilizing the parallel Fourier transform and simple multiplication operations. Thus, it takes less time than the state-of-the-art methods while maintaining the same performance, and it achieves higher performance than the existing methods within the same execution time, without the need for pre-estimating the number of inputs. Furthermore, the influence of the model mismatch has only a minor impact on the estimation performance, which allows for more efficient use of the spectrum resource in a distributed swarm scenario. Simulation results demonstrate the low complexity in sampling and computation, making it a more practical solution for real-time and distributed wideband spectrum sensing applications.
\end{abstract}

\begin{IEEEkeywords}
Genralized Coprime sampling, power spectrum sensing, non-sparsity, blind sensing, cyclostationary.
\end{IEEEkeywords}

\section{Introduction}
\IEEEPARstart{T}{he} demand for spectrum resources is increasing due to the rapid development of low-orbit satellite constellation systems (e.g. SpaceX, OneWeb), 5G/6G networks, and the Internet of Things (IoT), etc \cite{Chae_2023_Rethinking}-\cite{Zhou_2022_NewParadigm}. These applications are driving a growing demand for wideband spectrum sensing. Correspondingly, direct sampling requires a high-speed analog-to-digital converter (ADC) based on the Shannon-Nyquist sampling theorem, leading to a prohibitive volume of data and high energy cost.

Currently, the most widely used methods are sweep-tune sampling and filter band sampling \cite{Fang_2021_Recent}, both of which fall under the category of low-speed sampling. Nevertheless, the scanning scheme has a detection latency and may miss short-lived signals. Additionally, the filter band scheme has a complicated structure and is prone to serious channel crosstalk. As a result, there has been a trend towards using wideband spectrum sensing as a guide.

The recent compressive sensing (CS) theory provides a sub-sampling scheme with low-speed and large instantaneous bandwidth by utilizing the sparsity in the frequency domain \cite{Mishra_2017_Compressive}-\cite{Wu_2019_Deep}. The typical CS schemes include analog-to-information converter (AIC), multi-coset sampling (MCS), and multi-rate sampling (MRS). However, the sparse signal recovery algorithms have high computational complexity and are extremely sensitive to noise. On the other hand, the engineering implementation of these schemes also presents a major difficulty, restricting their application.

To overcome these difficulties, the approach to spectrum estimation has been shifted processing the original signal to analyzing its second-order statistics. In this context, the compressive covariance sensing (CCS) theory provides a wideband spectrum sensing scheme that operates at low-speed and has a large instantaneous bandwidth. It is reliable even in environments with a low signal-to-noise ratio (SNR) and non-sparse conditions. The typical CCS methods include entropy functions minimization \cite{Huang_2019_Sparse}, matrix norm minimization \cite{Jiang_2019_Wideband}, and Toeplize matrix completion \cite{Qin_2017_Generalized}, among others. However, all these methods are based on the reconstruction of the covariance matrix and the use of mutiple signal classification (MUSIC) algorithm, which have high computational complexity and require pre-estimation the number of signals.

Furthermore, a computationally efficient method is developed based on the relationship between the autocorrelation sequence and sub-sampling samples of the MCS scheme \cite{Yang_2020_Fast}. Building on this, a fast solution for generalized coprime sampling is introduced, which utilizes only parallel FFT and multiplication operations. As a result, it achieves reduced time and low estimation error, presenting a tradeoff between system performance and the number of degrees of freedom (DOFs). Moreover, model mismatch has little effect on performance, making a more practical solution for real-time and distributed wideband spectrum sensing applications.

Notations: The bold characters denote vectors. The notations $\mathbb{R}$, $\mathbb{N}$ and $\mathbb{N}^+$ represent the set of real numbers, nonnegative integers, and positive integers, respectively. The superscripts $(\cdot)^T$ and $(\cdot)^H$ indicate the transpose and conjugate transpose of a vector or a matrix, respectively. The operator $\circ$ signifies the Hadamard product, $|\cdot|^2$ signifies the element-wise square modulus of a vector, and ceil($\cdot$) signifies round up to an integer. The symbols $\boldsymbol{F}_{a}$ and $\boldsymbol{F}_{a}^{-1}$ mean the $a$-point fast Fourier transofrm (FFT) and the inverse fast Fourier transofrm (IFFT), respectively.

\section{Signal Model}
\begin{figure}[H]
	\centering
	\includegraphics[width=7.5cm]{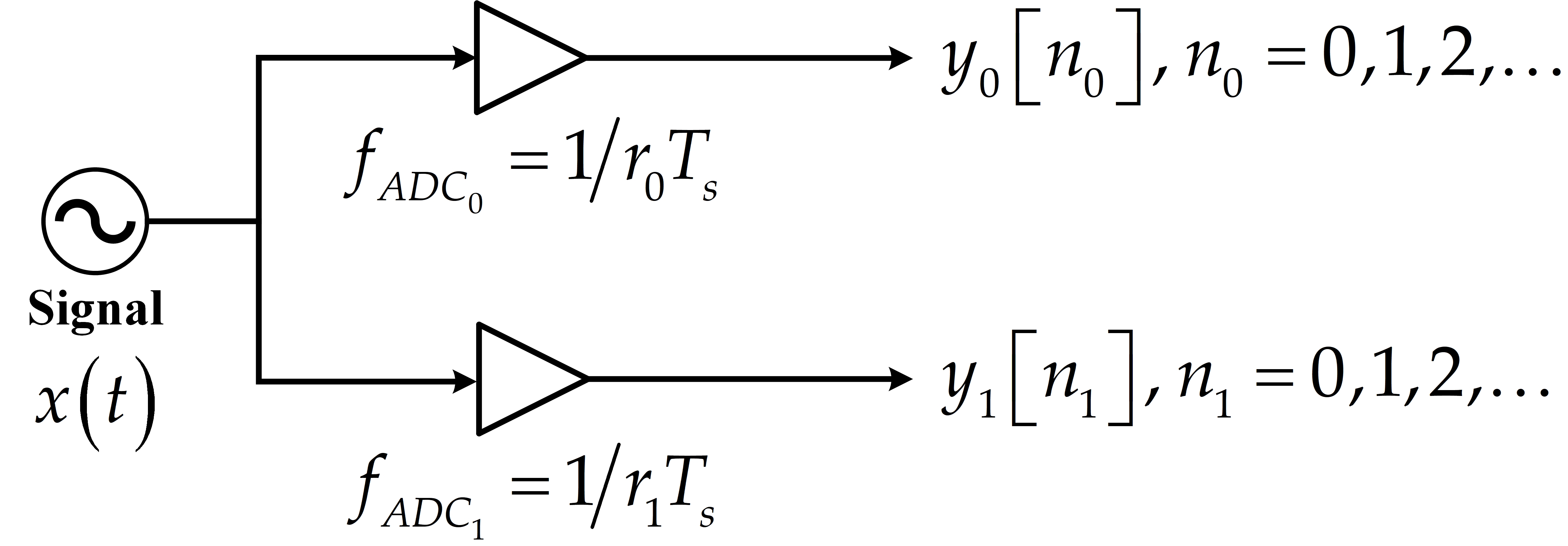}
	\caption{Coprime sampling scheme.}
	\label{fig_scheme}
\end{figure}

The generalized coprime sampling architecture consists of two uniform sub-Nyquist sampling channels, whose sampling periods are coprime multiples of the Nyquist sampling period. The introduction of two other operations, the multiple coprime unit factor $ p \in \mathbb{N}^+ $ and the non-overlapping factor $ q \in \mathbb{N}^+ $, increases the number of DOFs and improves the estimation accuracy. Therewith, the coprime sampling scheme is presented with sampling intervals $r_0 T_s$ and  $r_1 T_s$, as shown in Figure~\ref{fig_scheme}. Without loss of generality, it is assumed that $r_0 < r_1$ with $r_0, r_1 \in \mathbb{N}^{+}$, where $r_0$ and $r_1$ are coprime. And the sampling interval $T_s$ corresponds to the Nyquist sampling rate $f_s$.  

For a wide-sense stationarity or cyclostationary process $X(t), t \in \mathbb{R}$, it consists of a number of frequencies that are confined wihtin the bandwidth $B_s \leq f_s/2$. The outputs of two uniform sub-Nyquist sampling channels can then be expressed as
\begin{equation}
	\begin{split}
		y_0 [n_0] & = x [r_0 n_0] = X (r_0 n_0 T_s), \ n_0 \in \mathbb{N} \\
		y_1 [n_1] & = x [r_1 n_1] = X (r_1 n_1 T_s), \ n_1 \in \mathbb{N} 
	\end{split}
	\label{eq1}
\end{equation}
where $x[n], n \in \mathbb{N}$ denotes the Nyquist sampling samples, and the highest sampling rate of the coprime sampling system is given by $1/(r_0 T_s)=f_s/r_0$.

Accordingly, the elements of the sensing vector corresponding to the two coprime samplers can be donoted as 
\begin{equation}
	{a}_0 [i] = \left\{ \begin{array}{ll}
		1, & i = r_0 l_0 + k r_0 r_1\\
		0, & \textrm{elsewhere}
	\end{array}	\right.
	\label{eq2}
\end{equation}
and
\begin{equation}
	{a}_1 [i] = \left\{ \begin{array}{ll}
		1, & i = r_1 l_1 + (k +q) r_0 r_1\\
		0, & \textrm{elsewhere}
	\end{array}	\right.
	\label{eq3}
\end{equation}
where $ l_0 = 0,1,\dots,r_1 -1 $, $ l_1 = 0,1,\dots,r_0 -1 $, and $ k = 0,1,\dots,p-1 $.

From a data acquisition perspective, the output samples obtained from the generalized coprime sampling scheme are a subset of the Nyquist samples, positioned at
\begin{equation}
	\mathbb{P} = \{r_0 l_0 + k r_0 r_1\} \cup \{r_1 l_1 + (k +q) r_0 r_1\}
	\label{eq4}
\end{equation}

Based on the sensing vectors, the connection between the elements of the generalized coprime sampling vector and the Nyquist sampling vector can be expressed as
\begin{equation}
	{y} [n] = \left\{ \begin{array}{ll}
		x[n], & n \in \mathbb{P} \\
		0, & \textrm{elsewhere}
		\end{array}	\right.
	\label{eq5}
\end{equation}
and the elements of the connection sensing vector are defined as
\begin{equation}
	{a} [n] = \left\{ \begin{array}{ll}
		1, & n \in \mathbb{P} \\
		0, & \textrm{elsewhere}
	\end{array}	\right.
	\label{eq6}
\end{equation}

As a result, there is
\begin{equation}
	\mathbf{y} [n] = \mathbf{a} [n] \circ \mathbf{x} [n]
	\label{eq7}
\end{equation}
where $\mathbf{x} [n] = \big[ x[0], x[1], \dots, x[N-1] \big] ^T $, $\mathbf{a} [n] = \big[ a[0], a[1],$
$\dots, a[N-1] \big] ^T $ and $\mathbf{y} [n] = \big[ y[0], y[1], \dots, y[N-1] \big] ^T $ are all vectors of size $ N \times 1$. 

\section{Proposed Fast Solution}
Considering the widely-used unbiased estimation of the autocorrelation sequence for the output of generalized coprime sampling, the elements $r_y [m]$ can be given as
\begin{equation}
	r_y [m] = \frac{1}{N} \cdot \mathbf{y}[n] \mathbf{y}^H [n-m]
	\label{eq8}
\end{equation}

Substituting equation (\ref{eq7}) into equation (\ref{eq8}) yields
\begin{equation}
	\begin{aligned}
		r_y [m] 
		&= \frac{1}{N} \cdot (\mathbf{a}[n] \circ \mathbf{x}[n]) (\mathbf{a}^H [n-m] \circ \mathbf{x}^H [n-m])\\
		&= \frac{1}{N} \cdot \big( (\mathbf{a}[n] \mathbf{a}^H [n-m]) \circ (\mathbf{x}[n] \mathbf{x}^H [n-m]) \big) \\
		&= r_a [m] \circ r_x [m]
	\end{aligned}
	\label{eq9}
\end{equation}
where $\lvert m \rvert \leq N-1$. Thus, the power spectrum can be obtained by performing FFT on the autocorrelation sequence $ \{ r_x [m] \}$, which is derived from the autocorrelation sequence $\{ r_y [m] \}$ and $\{ r_a [m] \}$. Then, a practical solution that computationally efficient is employed to obtain the estimation of the autocorrelation sequence. The steps are as follows:

\vspace{6pt}
\textbf{Step 1:} Pad vectors $\mathbf{a} [n]$ and $\mathbf{y} [n]$ with an additional N-length zeros.
\begin{equation}
	{a}_{2N} [n] = \left\{ \begin{array}{ll}
		{a}_{N} [n], & 0 \leq n \leq N-1 \\
		0, & N \leq n \leq 2N-1
	\end{array}	\right.
	\label{eq10}
\end{equation}
and
\begin{equation}
	{y}_{2N} [n] = \left\{ \begin{array}{ll}
		{y}_{N} [n], & 0 \leq n \leq N-1 \\
		0, & N \leq n \leq 2N-1
	\end{array}	\right.
	\label{eq11}
\end{equation}

\vspace{6pt}
\textbf{Step 2:} Calculate the autocorrelation sequence based on the power spectrum estimation of vector $\mathbf{a}_{2N} [n] = \big[ a[0], a[1],$ 
$\dots, a[2N-1] \big] ^T$ and $\mathbf{y}_{2N} [n] = \big[ y[0], y[1], \dots, y[2N-1] \big] ^T$ by involving FFT and IFFT.
\vspace{3pt}
\begin{equation}
	\hat{\mathbf{r}}'_a [k] = \boldsymbol{F}_{2N}^{-1} \vert \boldsymbol{F}_{2N} \mathbf{a}_{2N} \vert ^2 /N
	\label{eq12}
\end{equation}
and
\begin{equation}
	\hat{\mathbf{r}}'_y [k] = \boldsymbol{F}_{2N}^{-1} \vert \boldsymbol{F}_{2N} \mathbf{y}_{2N} \vert ^2 /N
	\label{eq13}
\end{equation}
where $k=0,1,\dots,2N-1$.

\vspace{6pt}
\textbf{Step 3:} Truncate the autocorrelation sequence of interest according to the frequency resolution of the system $\Delta f$.
\begin{equation}
	\hat{r}_{y} [m] = \left\{ \begin{array}{ll}
		\hat{r}'_{y} [m], & 0 \leq m \leq M-1 \\
		\hat{r}'_{y} [m+2N], & -M+1 \leq m \leq -1
	\end{array}	\right.
	\label{eq14}
\end{equation}
and
\begin{equation}
	\hat{r}_{a} [m] = \left\{ \begin{array}{ll}
		\hat{r}'_{a} [m], & 0 \leq m \leq M-1 \\
		\hat{r}'_{a} [m+2N], & -M+1 \leq m \leq -1
	\end{array}	\right.
	\label{eq15}
\end{equation}
where $M = \textit{ceil} (fs/2/\Delta f) +1$. 

\vspace{6pt}
\textbf{Step 4:} Compute the autocorrelation sequence of the inputs using the obatined sequences $\{ \hat{r}_{y} [m] \}$ and $\{ \hat{r}_{a} [m] \}$.
\begin{equation}
	\hat{r}_{x} [m] = \hat{r}_{y} [m] ./ \hat{r}_{a} [m]
	\label{eq16}
\end{equation}
where $m=-M+1, \dots, -1,0,1, \dots, M-1$.

\vspace{6pt}
\textbf{Step 5:} Obtain the power spectrum estimation by taking the FFT of vector $\hat{\mathbf{r}}_x [m] = \big[  \hat{r}_{x} [-M+1] , \dots, \hat{r}_{x} [1], \hat{r}_{x} [0], \hat{r}_{x} [1], \dots, \hat{r}_{x} [M-1] \big] ^T$.
\begin{equation}
	\hat{\mathbf{S}}_x (\omega) = \vert \boldsymbol{F}_{2M-1} \hat{\mathbf{r}}_x [m] \vert
	\label{eq17}
\end{equation}

\begin{figure}[H]
	\centering
	\includegraphics[width=7cm]{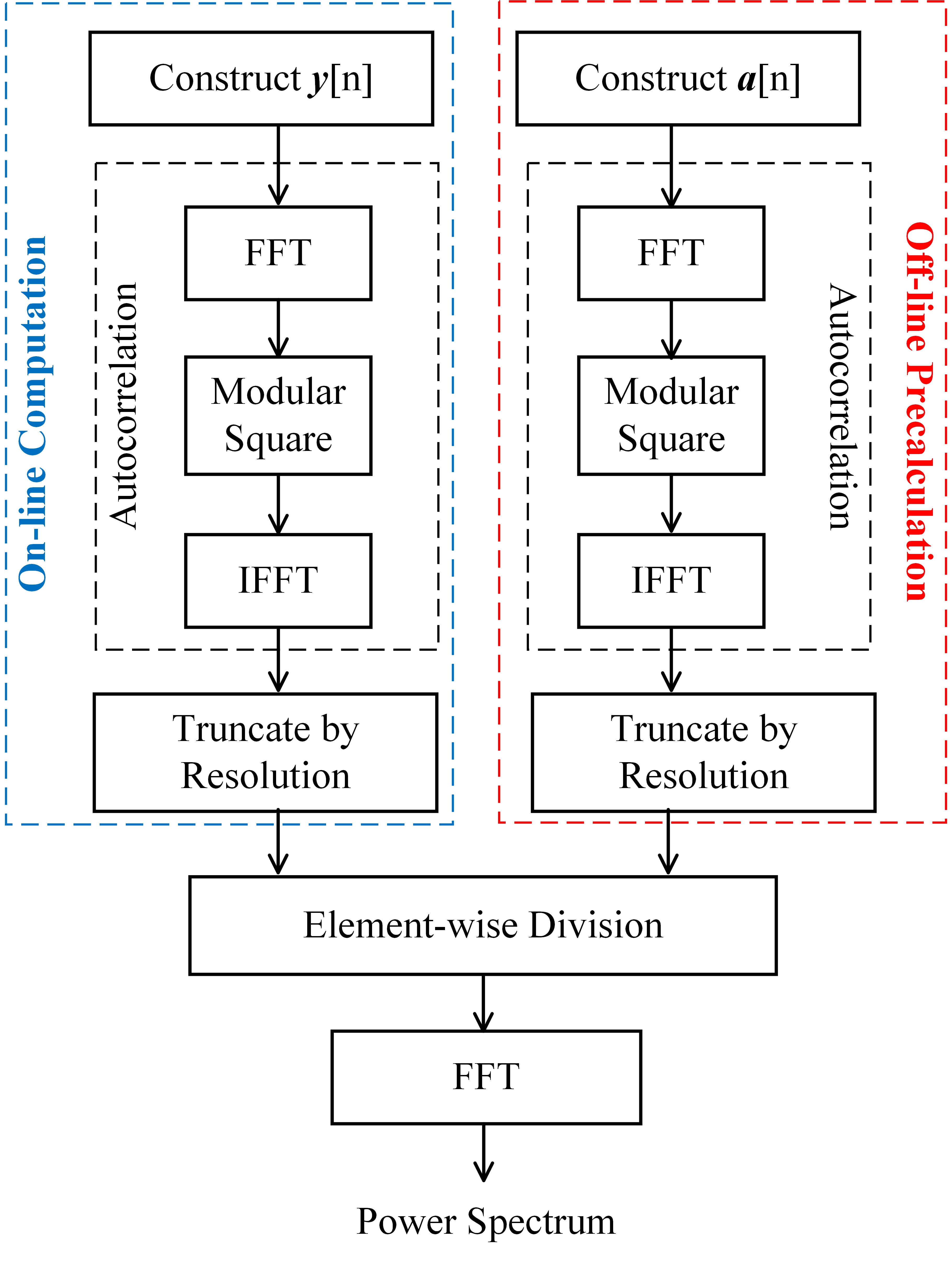}
	\caption{Block diagram of the proposed fast power spectrum sensing solution.}
	\label{fig_diagram}
\end{figure}

The block diagram illustrating the fast power spectrum sensing solution for generalized coprime sampling is depicted in Figure~\ref{fig_diagram}. The proposed solution is efficient, as it only involves FFT/IFFT operations and some basic multiplication operations. Moreover, the autocorrelation sequence $\{ \hat{r}_{a} [m] \}$ of the sensing vector can be pre-calculated offline, as shown in the red section of Figure~\ref{fig_diagram}. This calculation is solely dependent on the generalized coprime sampling scheme. 

Consequently, the computational complexity of the proposed solution involves the FFT of a $(2N)$-point sequence twice, resulting in $(2N) \text{log} (2N)$ floating-point operations according to (\ref{eq13}). In addition, the FFT of a $(2M-1)$-point sequence is executed once, leading to $(2M-1) \text{log} (2M-1)$ floating-point operations according to (\ref{eq17}). Upon incorporating $2M-1$ multiplication calculations, the total computational complexity necessitates $(4N) \text{log} (2N) + (2M-1) \text{log} (2M-1) + (2M-1)$ floating-point operations. This results in a lower computational complexity compared to the state-of-the-art methods. Meanwhile, it is feasible to compute the FFT operations in parallel effectively, which is a more suitable practical solution to meet the real-time wideband power spectrum sensing applications. 

\section{Simulation Results}
In the experiments, it is assume that there are $I$ inputs with identical powers, which are distributed in the frequency band $[2,18]$GHz. Subsequently, the coprime integers $r_0 = 3$, $r_1 = 4$ and the Nyquist sampling rate $f_s = 32$GHz are set. Furthermore, the relative root mean square error (RMSE) is adopted to evaluate the performance of the proposed fast power spectrum sensing method, which is defined as follows
\begin{equation}
	\text{Relative RMSE} (f_i) = \frac{1}{f_s} \sqrt{ \frac{1}{500 I} \sum_{j=1}^{500} \sum_{i=1}^I (\hat{f}_i(j) - f_i)^2}
	\label{eq18}
\end{equation}
where $\hat{f}_i(j)$ is the estimation of $f_i$ from the $j$th Monte Carlo trial, and five hundred Monte Carlo trials are conducted.

\begin{figure}[H]
	\centering
	\begin{minipage}{0.49\linewidth}
		\centering
		\includegraphics[width=0.9\linewidth]{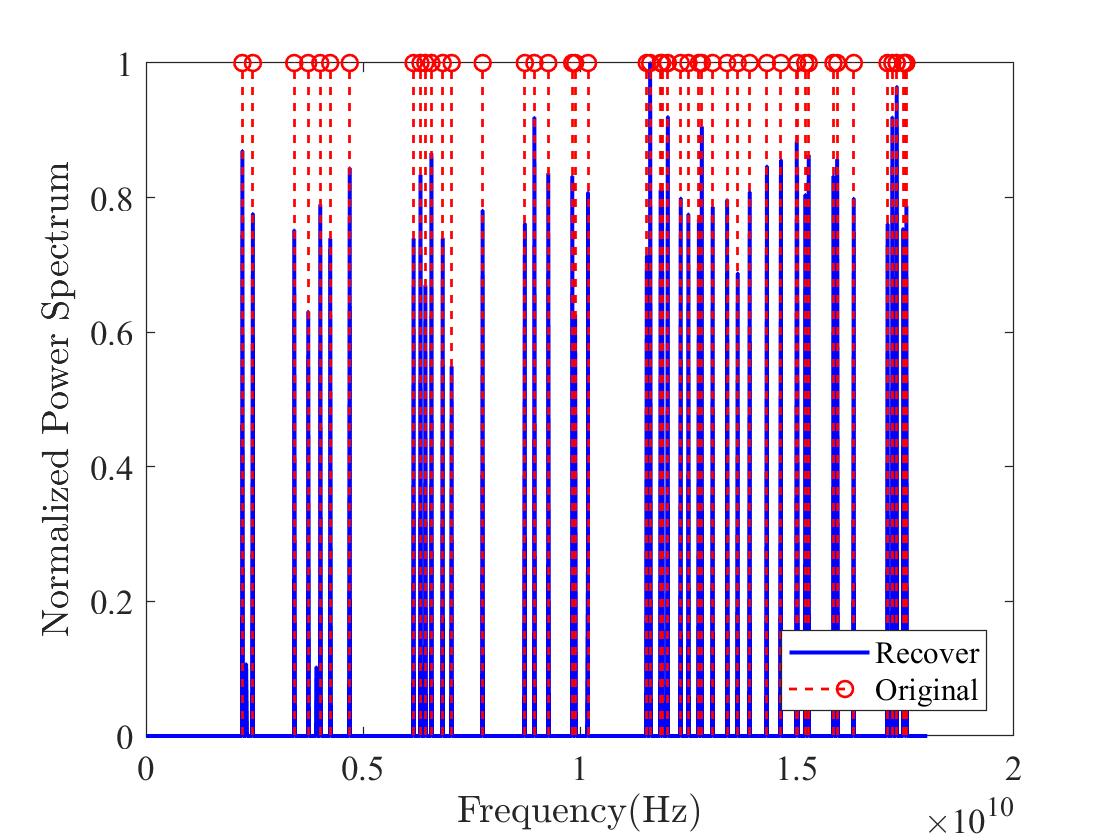}
		\text{(a) MP ($I=50$)}
		\label{fig_mp}
	\end{minipage}
	\begin{minipage}{0.49\linewidth}
		\centering
		\includegraphics[width=0.9\linewidth]{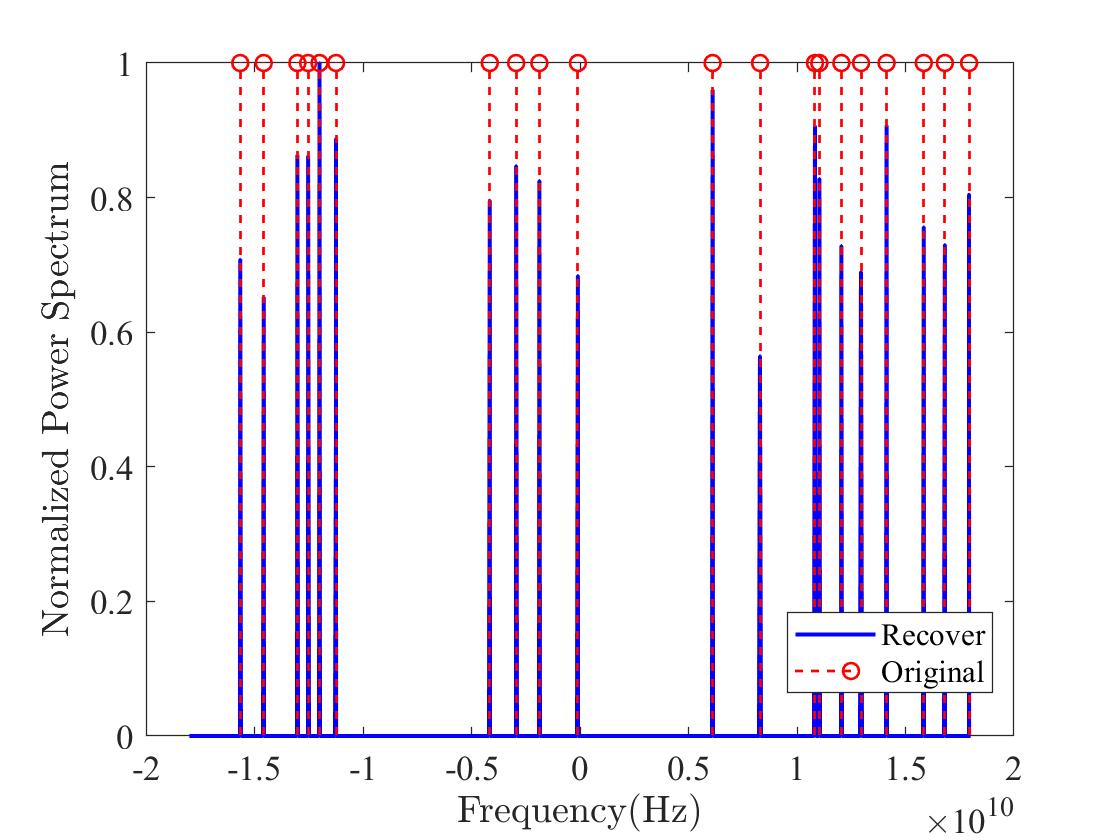}
		\text{(b) BPSK ($I=20$)}
		\label{fig_bpsk}
	\end{minipage}

	\begin{minipage}{0.49\linewidth}
		\centering
		\includegraphics[width=0.9\linewidth]{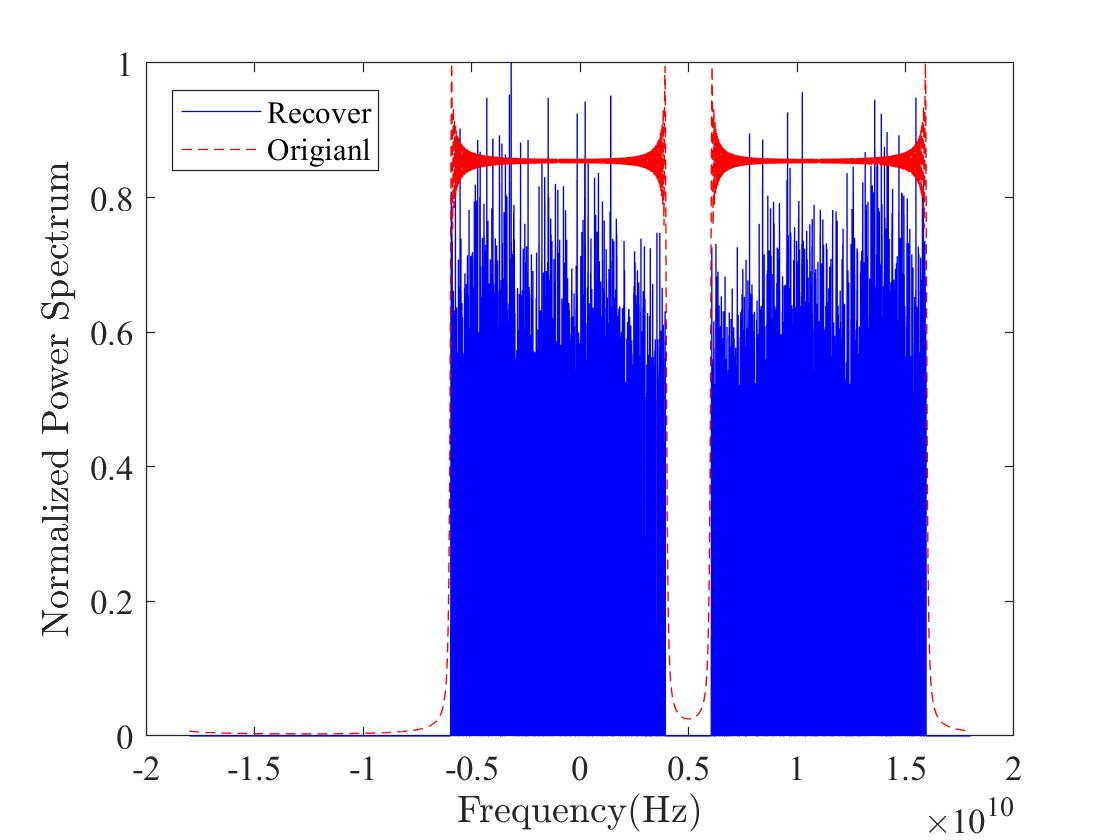}
		\text{(c) LFM ($I=2$)}
		\label{fig_lfm}
	\end{minipage}
	\begin{minipage}{0.49\linewidth}
		\centering
		\includegraphics[width=0.9\linewidth]{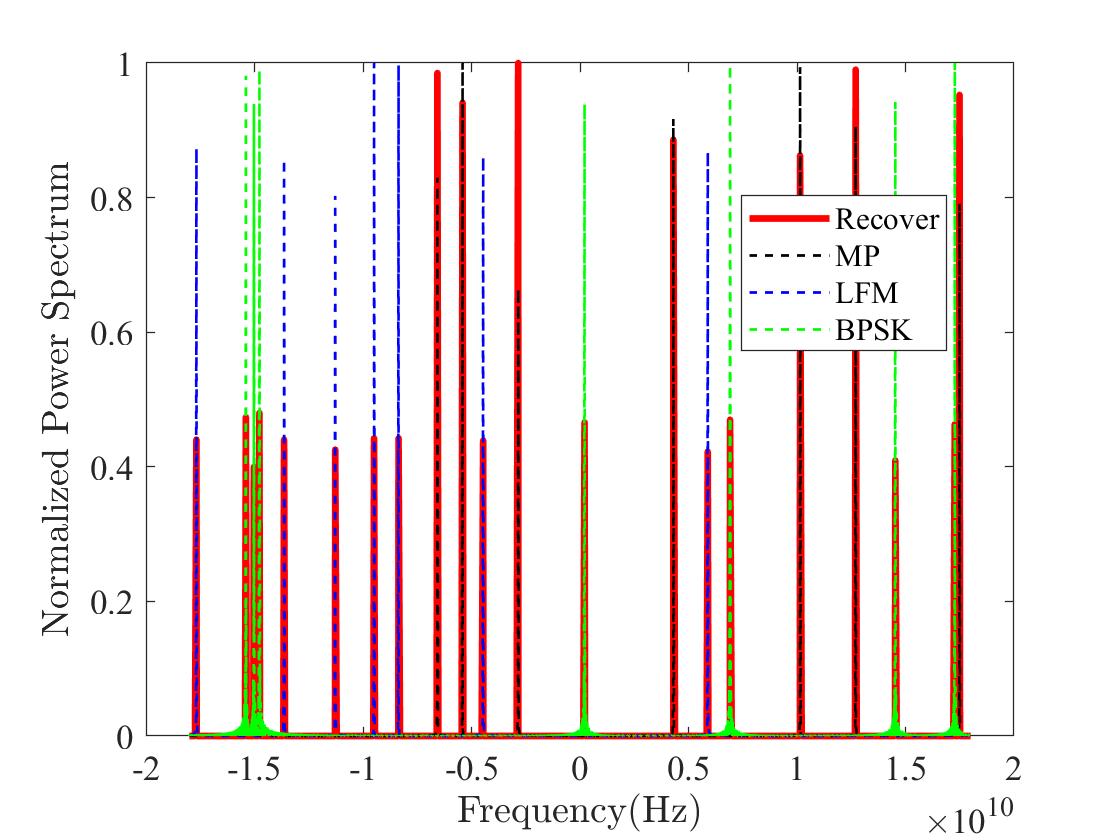}
		\text{(d) Mixed ($I=21$)}
		\label{fig_mix}
	\end{minipage}

	\caption{Estimated Power spectrum (SNR$=15$dB).}
	\label{fig_rec}
\end{figure}

Herein, the estimated power spectrum results are first displayed in Figure~\ref{fig_rec}, with $p=3000$ and an input SNR of 15dB. There are $I=50$ mono-frequency pulse (MP) signals randomly distributed in (a), $I=20$ binary phase shift keying (BPSK) signals for 1M symbols per second with random frequency and code depicted in (b), $I=2$ linear frequency modulation (LFM) signals with 10GHz bandwidth under $\pm 6$GHz initial frequencies given in (c) and $I=21$ mixture signals of three types for (d). As can be observed, all frequencies are estimated accurately with the proposed method. 

\begin{figure}[H]
	\centering
	\includegraphics[width=0.55\linewidth]{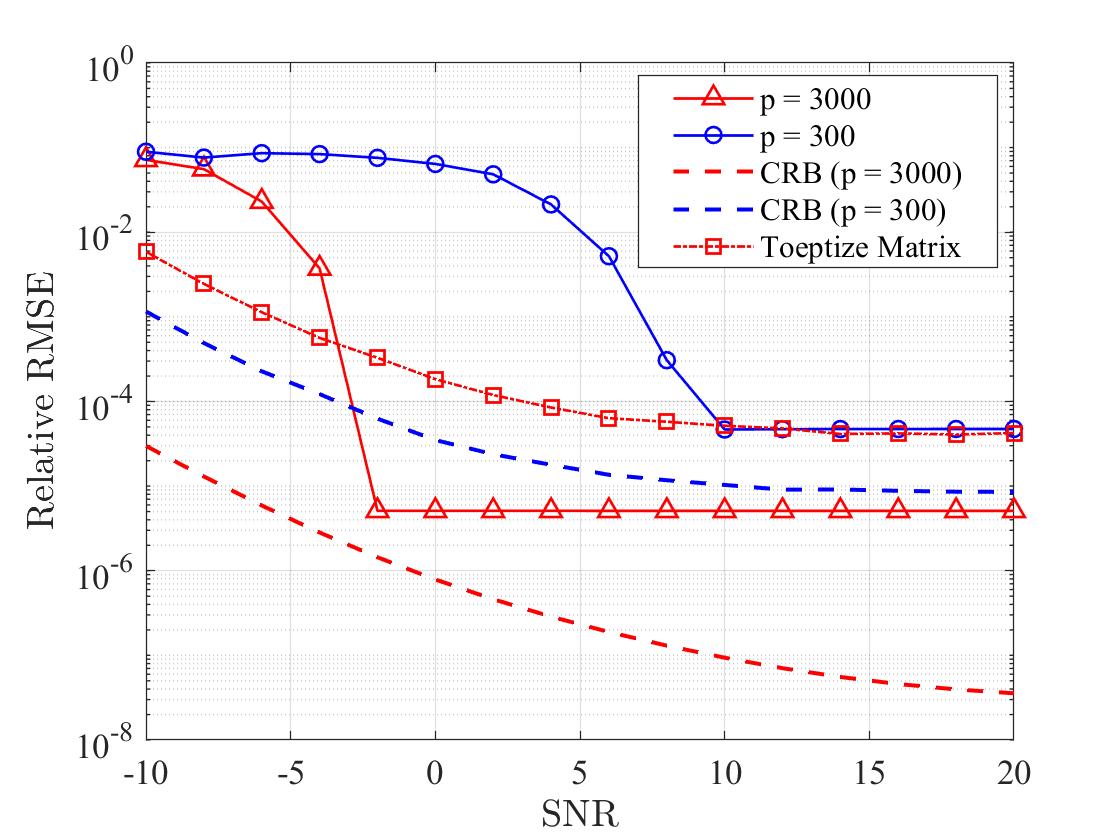}
	\caption{Relative RMSE versus SNR ($I=18$).}
	\label{fig_snr}
\end{figure}

\begin{figure}[H]
	\centering
	\includegraphics[width=0.55\linewidth]{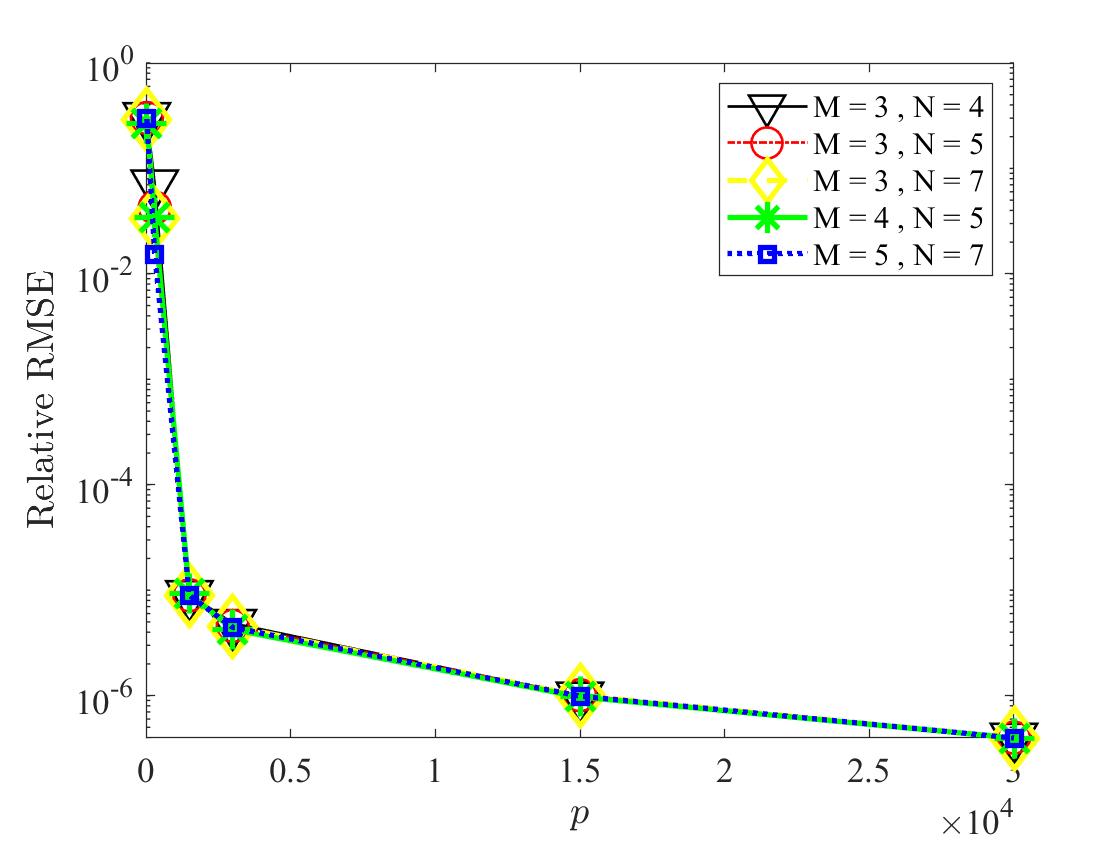}
	\caption{Relative RMSE versus $p$ ($I=18$, SNR$=0$dB).}
	\label{fig_fmin}
\end{figure}

As shown in Figure~\ref{fig_snr}, the RMSE results are compared as a function of the input SNR, where $I=18$ MP signals are used and the frequency is randomly selected. It can be seen that the RMSE tends to be stabilized when SNR is greater than -2dB for $p=300$. Under the same condition, the original method has the better performance at a low SNR by using the Toeplize matrix completion. Meanwhile, the estimation performance is improved as $p$ increases, due to that the DOF increases with $p$ and the resolution also improves. As expected, the Figure~\ref{fig_fmin} has the same result. However, the selection of coprime sampling rate makes less difference to performance when p is greater than 1000. That is because the system redundancy under simulation is enough.

\begin{figure}[H]
	\centering
	\includegraphics[width=0.55\linewidth]{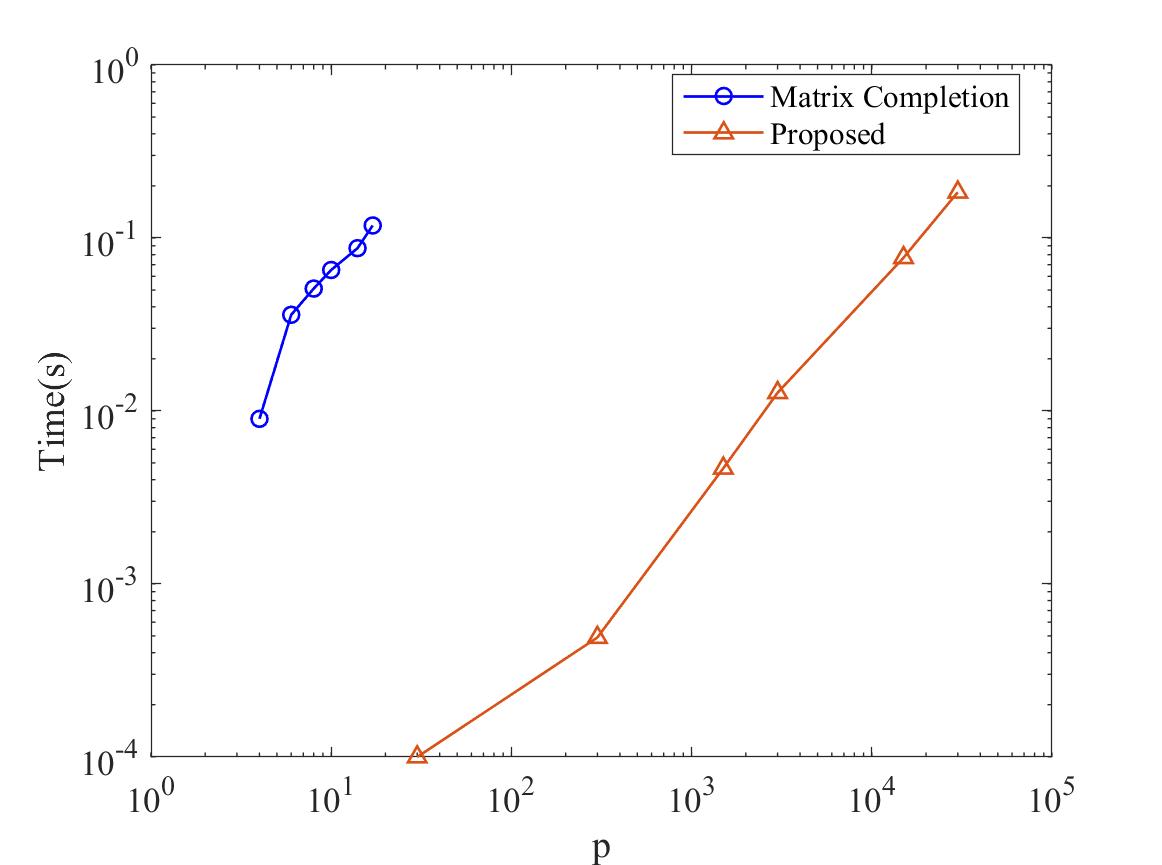}
	\caption{Execution time versus $p$ ($I=10$, SNR$=0$dB).}
	\label{fig_p}
\end{figure}

\begin{figure}[H]
	\centering
	\includegraphics[width=0.55\linewidth]{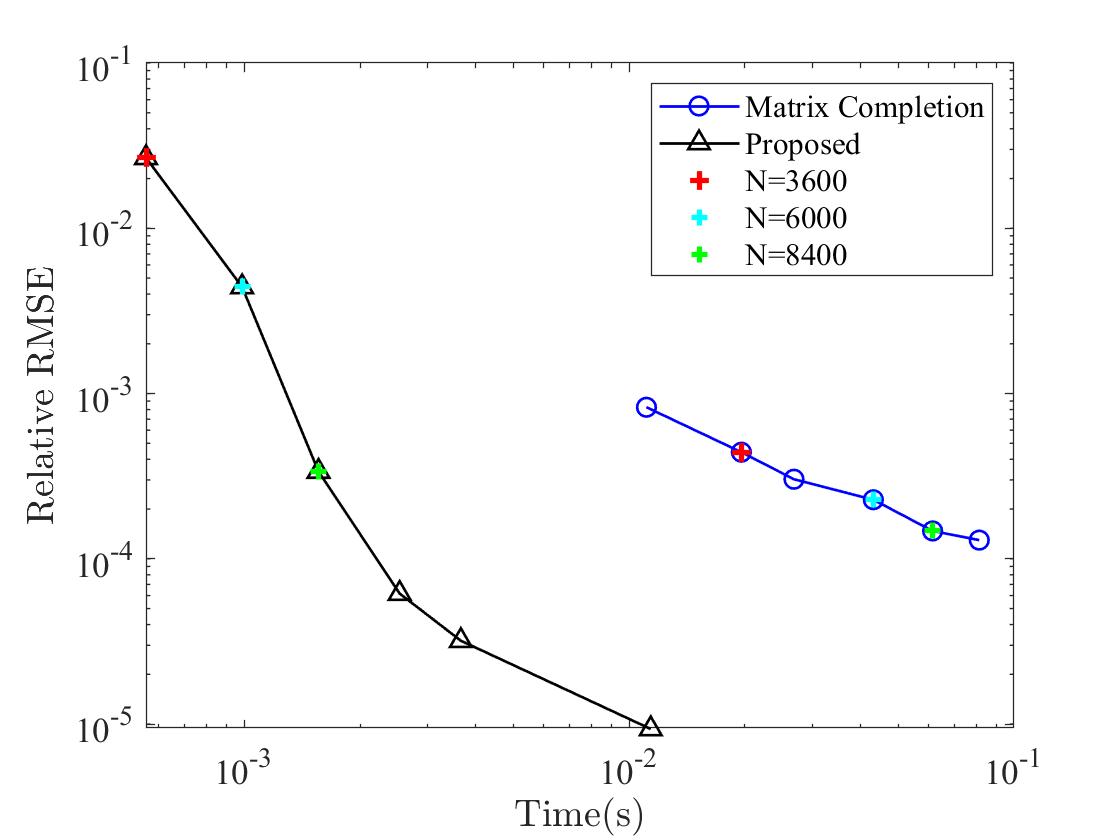}
	\caption{Execution time versus Relative RMSE under the same number of samples ($I=18$, SNR$=5$dB).}
	\label{fig_time}
\end{figure}

Furthermore, the multiple coprime unit factor $p$ not only affects the resolution, but also determines the execution time of algorithms. Thus, the execution time results are compared as a function of $p$ as displayed in Figure~\ref{fig_p}, where $I=10$ MP signals are used and the frequency is randomly selected with the 0dB SNR. It can be seen that the execution time of the proposed method has obvious advantages over the matrix completion method. Similar to the Figure~\ref{fig_time}, the proposed method outperforms the matrix completion method under the same execution time. Furthermore, the proposed method takes less time than the matrix completion method under the same performance. However, both methods rely on the larger number of samples. Therefore, there is a tradeoff between execution time and the system performance.

\begin{figure}[H]
	\centering
	\includegraphics[width=0.55\linewidth]{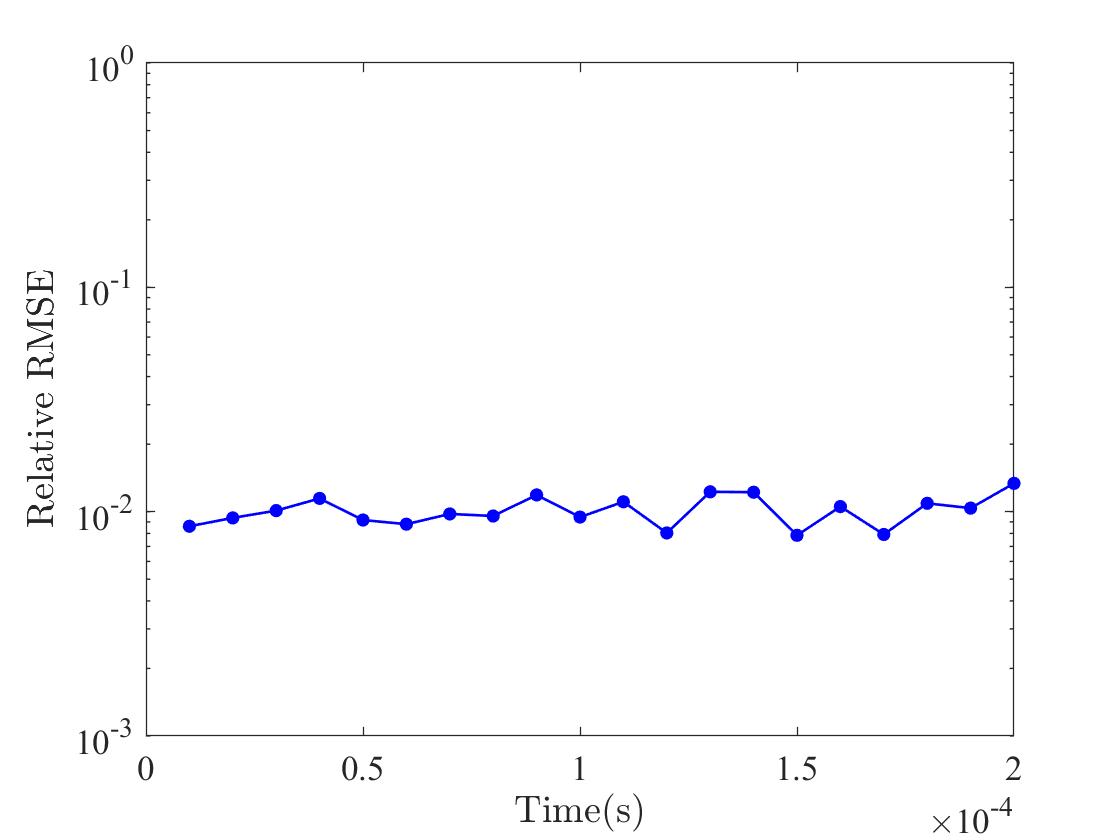}
	\caption{Relative RMSE versus model matching ($I=18$, SNR$=5$dB).}
	\label{fig_delay}
\end{figure}

At last, the influence of the model matching degree between the sensing vector and measurements on the performance is discussed in Figure~\ref{fig_delay} with the 5dB SNR. Here, $I=18$ frequencies are randomly selected for MP signals, and different time delay are used which is unknown for the sensing vector. As a result, the influence of the model mismatch has little fluctuation of RMSE within $200 \mu$s. That is interesting, which means that the fact potentially enables the spectrum resource in a distributed swarm scenario to be more efficiently utilized.

\section{Conclusions}
A fast power spectrum sensing solution for generalized coprime sampling is proposed, that only uses the parallel FFT and simple multiplication operations. It has obvious advantages over existing methods in terms of spectrum estimation performance and execution time, and there is no need to pre-estimate the number of inputs. Moreover, the influence of the model mismatch has little impact on the spectrum estimation performance, making it more suitable for further discussion on its application in the distributed swarm scenario.

\bibliographystyle{IEEEbib}
\bibliography{Refs_Manuscript}

\begin{thebibliography}{1}

\bibitem{Chae_2023_Rethinking}
Keunhong Chae, Jungin Park, and Yusung Kim,
\newblock ``Rethinking autocorrelation for deep spectrum sensing in cognitive
  radio networks,''
\newblock {\em IEEE Internet Things J.}, vol. 10, no. 1, pp. 31--41, 2023.

\bibitem{Zhou_2022_NewParadigm}
B.~Zhou, X.~Ma, T.~Kuang, and J.~Li,
\newblock ``New paradigm of electromagnetic spectrum space situation cognition:
  Spectrum semantic and spectrum behavior,''
\newblock {\em Journal of Data Acquisition and Processing}, vol. 37, no. 6, pp.
  1198--1207, 2022.

\bibitem{Fang_2021_Recent}
J.~Fang, B.~Wang, H.~Li, and Y.-C. Liang,
\newblock ``Recent advances on sub-nyquist sampling-based wideband spectrum
  sensing,''
\newblock {\em IEEE Wireless Commun.}, vol. 28, no. 3, pp. 115--121, 2021.

\bibitem{Mishra_2017_Compressive}
Amit~Kumar Mishra and Ryno~Strauss Verster,
\newblock {\em Compressive Sensing Based Algorithms for Electronic Defence},
\newblock Signals and Communication Technology. Springer International
  Publishing, 2017.

\bibitem{Wu_2019_Deep}
Yan Wu, Mihaela Rosca, and Timothy Lillicrap,
\newblock ``Deep compressed sensing,'' 2019-05-18.

\bibitem{Huang_2019_Sparse}
Shuai Huang and Trac~D Tran,
\newblock ``Sparse signal recovery via generalized entropy functions
  minimization,''
\newblock {\em Ieee Trans. Signal Process.}, vol. 67, no. 5, pp. 16, 2019.

\bibitem{Jiang_2019_Wideband}
K.~Jiang, Y.~Xiong, and B.~Tang,
\newblock ``Wideband spectrum sensing via derived correlation matrix completion
  based on generalized coprime sampling,''
\newblock {\em IEEE Access}, vol. 7, no. 1, pp. 117403--117410, 2019.

\bibitem{Qin_2017_Generalized}
S.~Qin, Y.~D. Zhang, M.~G. Amin, and A.~M. Zoubir,
\newblock ``Generalized coprime sampling of toeplitz matrices for spectrum
  estimation,''
\newblock {\em IEEE Trans. Signal Process.}, vol. 65, no. 1, pp. 81--94, 2017.

\bibitem{Yang_2020_Fast}
L.~Yang, J.~Fang, H.~Duan, and H.~Li,
\newblock ``Fast compressed power spectrum estimation: Toward a practical
  solution for wideband spectrum sensing,''
\newblock {\em IEEE Trans. Wireless Commun.}, vol. 19, no. 1, pp. 520--532,
  2020.

\end{thebibliography}

\vfill

\end{document}